\documentclass[12pt,paper]{JHEP3}
\usepackage{amssymb}
\usepackage{epsfig}
\def\del{\Delta}
\def\ddel{{}^\bullet\! \Delta}
\def\deld{\Delta^{\hskip -.5mm \bullet}}

\def\ddeld{{}^{\bullet}\! \Delta^{\hskip -.5mm \bullet}}

\newcommand{\be}{\begin{equation}}
\newcommand{\ee}{\end{equation}}
\newcommand{\bea}{\begin{eqnarray}}
\newcommand{\eea}{\end{eqnarray}}
\def\eqa{&=&}
\def\ccr{\nonumber\\}
\def\la{\langle}
\def\ra{\rangle}

\def\bchi{\bar{\chi}}
\def\bpsi{\bar{\psi}}
\def\bep{\bar{\epsilon}}
\def\ep{\epsilon}
\def\rldd{\rlap{\,/}\nabla}

\title{Worldline approach to vector and antisymmetric tensor fields}

\author{Fiorenzo Bastianelli\\
Dipartimento  di Fisica, Universit{\`a} di Bologna 
and  INFN, Sezione di Bologna, Via Irnerio 46, I-40126 Bologna, Italy \\ 
E-mail: \email{bastianelli@bo.infn.it}} 
\author{Paolo Benincasa\\
Department of Applied Mathematics, University of Western Ontario,
Middlesex College, London, ON, Canada N6A 5B7\\
E-mail: \email{pbeninca@uwo.ca}}
\author{ Simone Giombi\\
C.N. Yang Institute for Theoretical Physics,  
State University of New York at Stony Brook, 
Stony Brook, NY 11794-3840, USA\\
E-mail: \email{sgiombi@insti.physics.sunysb.edu}}

\abstract{
The $N=2$ spinning particle action describes the propagation of 
antisymmetric tensor fields, including vector fields as a special case. 
In this paper we study the path integral quantization on a one-dimensional 
torus of the $N=2$ spinning particle coupled to spacetime gravity.
The action has a local $N=2$ worldline supersymmetry with a gauged 
$U(1)$ symmetry that includes a Chern-Simons coupling. 
Its quantization on the torus produces the one-loop effective 
action for a single antisymmetric tensor. We use this worldline representation 
to calculate the first few Seeley-DeWitt coefficients for 
antisymmetric tensor fields of arbitrary rank in arbitrary dimensions.
As side results we obtain the correct trace anomaly of a spin 1 
particle in four dimensions as well as exact duality relations 
between differential form gauge fields.
This approach yields a drastic simplification over standard 
heat-kernel methods. It contains on top of the usual proper time 
a new modular parameter implementing the reduction to a single tensor field.
Worldline methods are generically simpler and more efficient 
in perturbative computations than standard QFT Feynman rules. 
This is particularly evident when the coupling to gravity is considered.}


\preprint{YITP-SB-05-10}

\begin{document}

\section{Introduction}
\label{sec:intro}

Worldline approaches often produce efficient tools for calculating
Feynman diagrams of standard quantum field theories,
see e.g. \cite{Schubert:2001he} for a review.
Recently, also gravitational interactions have been discussed 
in this framework by considering the path integral quantization of 
worldlines of particles of spin 0 and 1/2 embedded in a curved spacetime
\cite{Bastianelli:2002fv,Bastianelli:2002qw,Bastianelli:2003bg}.
This method has then been applied to study new processes 
\cite{Bastianelli:2004zp}. 
In this paper we wish to study the propagation of particles of spin 1
and, more generally, of antisymmetric tensor fields coupled to gravity.
The corresponding mechanical  model that must be quantized is the $N=2$ 
spinning particle discussed in \cite{Brink:1976uf},
and further analyzed and extended in 
\cite{Berezin:1976eg,Gershun:1979fb,Howe:1988ft,Howe:1989vn,Marcus:1994mm}.
This model with a suitable Chern-Simons coupling 
is known to describe antisymmetric tensor fields of arbitrary 
rank \cite{Howe:1989vn}.
Our aim is to describe the one-loop effective action of 
a spin 1 particle and, more generally, of antisymmetric tensor fields 
in terms of worldline path integrals.
Indeed this is possible: proceeding with the quantization
of the spinning particle one obtains a quite interesting 
representation of the one-loop effective action.
This effective action is written in terms of
an integration over two moduli: the standard proper time and 
a new parameter related to the gauge fixing of the $U(1)$ symmetry.
The effect of this new modular parameter is to restrict 
the propagation to the sector which corresponds to a single tensor field.
The worldline representation produces a drastic simplification
over standard heat-kernel methods, and gives us the chance of
computing the first few Seeley-DeWitt coefficients for antisymmetric 
tensor fields of {\em arbitrary rank} in arbitrary dimensions. 
Previous studies for a worldline description
of particles of spin 1 were presented in 
\cite{Strassler:1992zr,Reuter:1996zm}, where
the projection to the physical states of the spin 1 particle 
was achieved by using a certain limiting procedure.
It differs from the construction presented here. 
However, we do not consider the couplings to 
background electromagnetism or Yang-Mills fields which, on the other hand, 
were investigated in \cite{Strassler:1992zr,Reuter:1996zm}.

The paper is organized as follows. In section 2 we review the 
$N=2$ spinning particle and remind that it describes the propagation
of a gauge potential $p$-form $A_p$ with standard gauge invariant 
Maxwell action. It turns out that the particle description is achieved 
directly in terms of the field strength, the $(p+1)$-form
$F_{p+1}=dA_p$.
In section 3 we consider the path integral quantization of the $N=2$
spinning particle action on the torus, and describe its gauge fixing.
In particular,  the  gauge fixing of the $U(1)$ symmetry produces 
a new modular parameter on top of the usual proper time.
There is no need of summing over the spin structures of the fermions
as the integration over the $U(1)$ modulus effectively interpolates 
between all boundary conditions. 
The gauge fixed path integral thus obtained gives a novel representation of 
the one-loop effective action for a $p$-form.
In section 4 we discuss the perturbative evaluation of this
effective action using an expansion in the proper time.
This way we are able to compute the first few Seeley-DeWitt coefficients
for a $p$-form in arbitrary dimensions, namely 
the coefficients $ a_0 $, $ a_1 $ and $ a_2 $.
As a side result we obtain the trace anomaly for a 1-form, 
i.e. a spin 1 particle, in 4 dimensions using worldline methods.  
In section 5 we derive exact duality relations between
differential forms, and then present our conclusions.
In the appendix we collect some technical results on the worldline 
propagators and determinants, and on the dimensional regularization
of the $N=2$ nonlinear sigma model.

\section{A brief review of the $N=2$ spinning particle}

The $N=2$ spinning particle action is characterized by a $N=2$ extended 
supergravity on the worldline. The gauge fields $(e, \chi, \bar \chi, a)$
of the $N=2$ supergravity contain in particular 
the einbein $e$ which gauges worldline translations, complex 
conjugate gravitinos $\chi$ and $\bchi$ which gauge the $N=2$ worldline
supersymmetry, and a standard gauge field  $a$ for the $U(1)$ 
symmetry which rotates by a phase the worldline fermions and gravitinos. 
The einbein and the gravitinos correspond to constraints that eliminate 
negative norm states and make the particle model consistent with unitarity.
The constraints arising from the gauge field $a$ makes the model 
irreducible, eliminating some further degrees of freedom
\cite{Gershun:1979fb,Howe:1988ft,Howe:1989vn}.

The action in flat target spacetime is most easily deduced by starting with 
a model with $N=2$ extended  rigid supersymmetry, and then gauging 
its symmetries. The rigid model is described in a graded phase space 
with real bosonic variables $(x^\mu,p_\mu)$ and complex fermionic 
variables $(\psi^\mu, \bpsi_\mu)$ 
($\bpsi_\mu$ is the complex conjugate of $\psi_\mu$).
It is given by the following real action
\be
S=\int dt \Big [ p_\mu \dot x^\mu + i \bpsi{}_\mu \dot \psi^\mu 
-\frac{1}{2} \eta_{\mu\nu} p^\mu p^\nu \Big ]  
\label{real-action}
\ee
where the indices $\mu , \nu =0,1,\cdots,D-1$ are spacetime indices
and $\eta_{\mu\nu}\sim (-,+,\cdots,+)$ is the Lorentz metric
used to lower and raise spacetime indices.
A dot denotes as usual the time derivative.
The graded Poisson brackets are then given by 
$\{x^\mu, p_\nu \}_{PB}=\delta^\mu_\nu$ and  
$\{\psi^\mu,\bar \psi_\nu\}_{PB} = -i \delta^\mu_\nu $.
This action  is manifestly Poincar\'e invariant in target space 
and thus describes a relativistic model which, however, is not unitary 
at this stage.
The lack of unitarity is due to negative norm states which appear 
in the Hilbert space because of the timelike value the index 
$\mu$  can take on the bosonic and fermionic variables.
As well-known in relativistic string and particle theory, unitarity 
can be recovered by imposing suitable constraints. This can be
achieved as follows. The action (\ref{real-action})
has on the worldline a rigid $N=2$ supersymmetry 
generated by the charges
\bea
H = {1\over 2} p_\mu p^\mu \ , \quad
Q = p_\mu \psi^\mu \ , \quad
\bar Q = p_\mu \bar \psi^\mu \ , \quad
J =  \bar \psi^\mu \psi_\mu  \ .
\eea
The whole symmetry algebra can be gauged since the charges 
close under Poisson brackets and can be considered as a set of first class 
constraints
\bea
\{Q,\bar Q\}_{PB} = -2i H \ , \quad
\{J,Q\}_{PB} = i Q  \ , \quad
\{J,\bar Q\}_{PB} = - i \bar Q  
\label{algebra}
\eea
(other Poisson brackets  vanish). These constraints are enough to recover 
unitarity, as it will be evident.
Introducing the gauge fields $G=(e, \bar \chi, \chi, a)$ which correspond
to the constraints $C=(H,Q,\bar Q, J)$  gives the action
\bea
S \eqa \int dt 
\Big [ p_\mu \dot x^\mu + i  \bpsi_\mu \dot \psi^\mu 
- e H - i \bchi Q- i \chi \bar Q -  a J \Big ]  \ccr
\eqa 
\int dt 
\Big [ p_\mu \dot x^\mu + i \bpsi_\mu \dot \psi^\mu 
-{1\over 2} e p_\mu p^\mu - i \bchi p_\mu \psi^\mu 
- i \chi p_\mu \bpsi^\mu - a \bpsi^\mu \psi_\mu 
\Big ]  \ .
\eea
The gauge transformations on the phase space variables 
are generated through Poisson brackets by
the generator 
$G \equiv \xi H + i\bep Q + i\epsilon \bar Q + \alpha J$, where 
$\xi, \bep , \epsilon, \alpha $  are local parameters with appropriate 
Grassmann parity,
\bea
\delta x^\mu \eqa \{ x^\mu , G \}_{PB} 
= \xi p^\mu + i\bep \psi^\mu  + i\ep \bpsi^\mu  
\ccr
\delta p_\mu \eqa \{ p_\mu , G \}_{PB} 
= 0 \ccr
\delta \psi^\mu \eqa \{ \psi^\mu , G \}_{PB} 
= -\epsilon p^\mu - i \alpha  \psi^\mu \ccr
\delta \bpsi^\mu \eqa \{ \bpsi^\mu , G \}_{PB} 
= -\bep p^\mu +i \alpha  \bpsi^\mu 
\label{gt1}
\eea
while on gauge fields the gauge transformations are easily obtained with 
the help of the constraint algebra (\ref{algebra})
\bea
\delta e \eqa \dot \xi + 2 i \bchi \ep+ 2 i \chi \bep \ccr
\delta \chi \eqa \dot \ep  +i a \epsilon - i \alpha   \chi \ccr
\delta \bchi \eqa \dot {\bep} - i a \bep + i \alpha  \bchi \ccr
\delta a \eqa \dot \alpha
\ .
\label{s2-gauge-tr}
\eea
Eliminating algebraically the momenta $p_\mu$ by using their equations 
of motion 
\be
p^\mu = {1\over e}(\dot x^\mu -i \bchi\psi^\mu-i \chi\bpsi^\mu)
\label{s2-peom}
\ee
one obtains the action in configuration space
\be
S=\int dt 
\Big [ {1\over 2} e^{-1} (\dot x^\mu -i \bchi\psi^\mu-i \chi\bpsi^\mu)^2
+ i  \bpsi_\mu \dot \psi^\mu -  a \bpsi^\mu \psi_\mu \Big ]  \ .
\ee
The corresponding gauge invariances can be easily deduced from the phase
space ones using (\ref{s2-peom}). 

In addition, one can add a Chern-Simons term for the gauge field $a$ 
\be
S_{CS}= q \int dt\, a 
\ee
which is obviously invariant under the gauge transformations 
(\ref{s2-gauge-tr}).
Absence of anomalies requires a quantization of the Chern-Simons 
coupling \cite{Elitzur:1985xj}
\be
q= {D\over 2} - p -1 \ , \quad \quad p\ {\rm integer}.
\ee
With this precise coupling the $N=2$ spinning particle describes an 
antisymmetric gauge field of rank $p$ (and corresponding field strength 
of rank $p+1$). In fact the gauge field $a$ with this Chern-Simons coupling 
produces the constraint  $J={D\over 2} - p -1$ instead of $J=0$.
A vector field (spin 1) has $p=1$ and thus does not need a Chern-Simons 
coupling in $D=4$, though such a term will be needed
in different dimensions.

Let us derive some of these statements by briefly
reviewing the canonical quantization of the model.
The phase space variables are turned into operators
satisfying the following (anti)commutation relations (we use $\hbar=1$)
\be
[ \hat x^\mu , \hat p_\nu ] = i \delta^\mu_\nu \ ,
\quad \quad
\{ \hat \psi^\mu, \hat \psi^\dagger_\nu \} = \delta^\mu_\nu \ . 
\ee
States of the full Hilbert space can be described by functions
of the coordinates $x^\mu$ and $\psi^\mu$. By 
$x^\mu$ we denote the eigenvalues of the operator $\hat x^\mu$, while
for the fermionic variables we use bra coherent states
defined by
\be
\la \psi | \hat \psi^\mu = \la \psi | \psi^\mu \ = \psi^\mu  \la \psi | \ .
\ee 
Any state $ |\phi\ra $ can then be described by the wave function
\be 
\phi(x,\psi)\equiv (\la x|\otimes \la \psi| ) |\phi\ra 
\ee
and since the $\psi$'s are Grassmann variables the wave function has the 
following general expansion
\be
\phi(x,\psi)
= F(x) + F_\mu(x) \psi^\mu + {1\over 2}
F_{\mu_1\mu_2}(x) \psi^{\mu_1}\psi^{\mu_2} +\ldots +
{1\over D!} F_{\mu_1\ldots\mu_D}(x) \psi^{\mu_1}\cdots \psi^{\mu_D} \ .
\label{expan}
\ee
The classical constraints $C$ now become operators $\hat C$  
which are used to select the physical states through the requirement
$\hat C | \phi_{phys}\ra=0$. 
In the above representation they take the form of differential operators
\bea 
\hat H = -{1\over 2}\partial_\mu \partial^\mu \ , \ \
\hat Q = -i \psi^\mu \partial_\mu \ , \ \
\hat Q^\dagger = -i \partial_\mu {\partial \over\partial \psi_\mu} \ , \ \
\hat J = - {1\over 2}\Big [\psi^\mu, {\partial \over\partial \psi^\mu}
 \Big ] - q \quad
\eea
where we have redefined $\hat J$ to include the Chern-Simons coupling
and antisymmetrized $\hat \psi^\mu$ and $\hat \psi^\dagger_\mu$
to resolve an ordering ambiguity.
The constraint $\hat J| \phi_{phys}\ra=0$ selects states with only 
$p+1$ $\psi$'s (recall that $q \equiv {D\over 2} - p -1 $), namely
\be
\phi_{phys} (x,\psi) =
{1\over (p+1)!} F_{\mu_1\ldots\mu_{p+1}}(x) \psi^{\mu_1}\cdots 
\psi^{\mu_{p+1}}  \ .
\ee
The constraints $\hat Q| \phi_{phys}\ra = 0$ gives the Bianchi identities
\be
\partial_{[\mu} F_{\mu_1\ldots\mu_{p+1}]}(x) =0
\ee
and the constraint $\hat Q^\dagger| \phi_{phys}\ra =0 $ produces the Maxwell 
equations
\be
\partial^{\mu_1} F_{\mu_1\ldots\mu_{p+1}}(x) =0 \ .
\ee
The constraint $\hat H| \phi_{phys}\ra =0 $ is
automatically satisfied as a consequence of the algebra 
$ \{ \hat Q , \hat Q^\dagger \} = 2 \hat H $. 

Thus we see that the $N=2$ spinning particle describes 
the propagation of a standard $p$-form gauge potential  
$A_{\mu_1\ldots\mu_{p}}$ in a gauge invariant way, namely through its
$F_{\mu_1\ldots\mu_{p+1}}$  field strength. 
The path integral approach of this model has also been
used to obtain the free propagator for a tensor field,
thus confirming the physical spectrum just discussed 
\cite{Papadopoulos:1989rg,Pierri:1990rp,Rivelles:1990dq,Marnelius:1993ba}.

Finally, we review the coupling to spacetime gravity. 
This coupling can be achieved by 
suitably covariantizing the constraints $H,Q,\bar Q, J$.
It is convenient, though not necessary, to use flat indices
for the worldline fermions by introducing the vielbein $e_\mu{}^a $ 
and the corresponding spin connection $\omega_\mu{}^{ab}$.
The action reads still as 
\bea
S \eqa \int dt 
\Big [ p_\mu \dot x^\mu + i  \bpsi_a \dot \psi^a 
- e H - i \bchi Q- i \chi \bar Q -  a J \Big ]  
\eea
but with covariantized constraints
(we now include the Chern-Simons term in $J$) 
\bea
J \eqa  \bar \psi^a \psi_a - q \ccr
Q \eqa  \psi^a e_a{}^\mu \pi_\mu \ccr
\bar Q \eqa  \bar \psi^a e_a{}^\mu \pi_\mu \ccr
H \eqa {1\over 2} g^{\mu\nu}\pi_\mu \pi_\nu 
- {1\over 2} R_{abcd} \bar \psi^a \psi^b \bar \psi^c \psi^d \ .
\eea
Here we have defined the ``covariant'' momentum
\be
\pi_\mu \equiv p_\mu - i \omega_{\mu ab} \bar \psi^a \psi^b 
\ee
which becomes the Lorentz covariant derivative upon canonical quantization.
The covariantizations of $Q$ and $\bar Q$ are  easy to guess.
Then one may use the algebra to identify $H$. Of course 
one must also check that the full constraint algebra remains unchanged.
For example, $\{Q ,Q\}_{PB} =0 $  is verified using 
the cyclic identity satisfied by the Riemann tensor.
Elimination of the momentum $p_\mu$ gives the 
configuration space action 
\bea
S \eqa \int dt 
\Big [ {1\over 2} e^{-1} g_{\mu\nu}
(\dot x^\mu -i \bchi\psi^\mu-i \chi\bpsi^\mu)
(\dot x^\nu -i \bchi\psi^\nu-i \chi\bpsi^\nu) \ccr
&+& i  \bpsi_a (\dot \psi^a + \dot x^\mu \omega_\mu{}^a{}_b \psi^b
+i a \psi^a )
+{e\over 2} R_{abcd} \bar\psi^a \psi^b \bar \psi^c \psi^d 
+ q a  \Big ]  \ .
\eea
This is the action we are going to quantize on the torus in the next sections.
Actually, for simplicity,  we prefer to use euclidean conventions. 
So we perform a Wick rotation to euclidean time 
($t\to -i \tau$, and also $ a\to i a$ to keep the gauge group $U(1)$ compact)
which produces the euclidean action ($ S_E=-i S $)
 \bea
S_E \eqa \int_0^1 d\tau 
\Big [ {1\over 2} e^{-1} g_{\mu\nu}
(\dot x^\mu - \bchi\psi^\mu- \chi\bpsi^\mu)
(\dot x^\nu - \bchi\psi^\nu- \chi\bpsi^\nu) \ccr
&+&  \bpsi_a (\dot \psi^a + \dot x^\mu \omega_\mu{}^a{}_b \psi^b
+ i a \psi^a )
-{e\over 2} R_{abcd} \bar\psi^a \psi^b \bar \psi^c \psi^d 
- i q a  \Big ] 
\label{full-conf-action}
\eea
where $\tau\in [0,1]$ parametrizes  the torus.
From now on we will drop the subscript on $S_E$ as no confusion 
should arise. Before closing this section, we list the gauge 
transformations of the supergravity multiplet in euclidean time,
as they will be needed to study the gauge fixing 
\bea
\delta e \eqa \dot \xi + 2  \bchi \ep+ 2  \chi \bep \ccr
\delta \chi \eqa \dot \ep  + i a \epsilon - i \alpha   \chi \ccr
\delta \bchi \eqa \dot {\bep} -  i a  \bep + i \alpha  \bchi \ccr
\delta a \eqa \dot \alpha
\ .
\label{24}
\eea

\section{Quantization on a torus}

We have seen that the action  (\ref{full-conf-action}) describes the 
propagation of a $p$-form in a background metric $g_{\mu\nu}$, 
or vielbein $e_\mu{}^a$.

\[
\raisebox{-1cm}
{\includegraphics*[112pt,678pt][227pt,730pt]{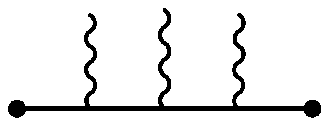}}
\]
\begin{center}
{\bf Figure 1:} Propagation of a $p$-form, wavy lines represent external 
gravitons
\end{center}

\noindent
Its quantization on a torus is then expected 
to produce the one-loop effective action
$\Gamma_{p}^{QFT} [g_{\mu\nu}]$
due to the virtual propagation of a $p$-form gauge field
in a gravitational background 
\be
\Gamma^{QFT}_{p}[g_{\mu\nu}]
\sim Z[g_{\mu\nu}] = \int_{T^1} {{\cal D} G {\cal D} X 
\over {\rm Vol(Gauge)}} \, e^{-S[X, G;g_{\mu\nu}]}
\label{25}
\ee
where $G=(e, \chi, \bar \chi,a)$ and $X=(x^\mu,\psi^\mu,\bar \psi^\mu)$
indicate the dynamical fields that must be integrated over,
and  $S[X, G;g_{\mu\nu}]$ denotes the action in (\ref{full-conf-action}).
Division by the volume of the gauge group reminds of the 
necessity of fixing the gauge symmetries.

\[
\raisebox{-1cm}
{\includegraphics*[266pt,654pt][348pt,732pt]{graphs.epsi}}
\]
\begin{center}
{\bf Figure 2:} Loop of a $p$-form in a gravitational background
\end{center}

\noindent
The torus is described by taking the parameter $\tau \in [0,1]$
and imposing periodic boundary conditions on the bosonic fields
$x^\mu$ and $e$ (the gauge field $a$ is instead treated as a connection).
As for the fermions we take antiperiodic boundary conditions,
and we shall soon understand why this is sufficient.
The gauge symmetries can be used to fix the supergravity 
multiplet to $\hat G=(\beta, 0,0,\phi)$, where $\beta$ and $\phi$ 
are the leftover bosonic moduli that must be integrated over.
The parameter $\beta$ is the usual proper time
\cite{Henneaux:1982ma,Cohen:1985sm,Polyakov}, 
while the parameter $\phi$ is a phase that corresponds to the only 
modular parameter that the gauge field $a$ can have on the torus.
Note that the gravitinos $\chi$ and $\bar\chi$ are antiperiodic and 
can be completely gauged away using (\ref{24}), leaving no moduli.

It is worthwhile to discuss more extensively  how the modular parameter 
$\phi$ arises. The action for the fermions in (\ref{full-conf-action})
is of the standard form (the target space geometry is inessential for 
this particular gauge fixing, and one can take it flat)
\be
S \sim \int_0^1 d\tau\, \bar \psi (\dot \psi +ia \psi) \ .
\label{act-fer}
\ee
Finite gauge transformations are given by
\bea
\psi (\tau) \ &\to& \ \psi'(\tau) = e^{-i \alpha (\tau)} \psi (\tau) \cr
\bar \psi (\tau) \ &\to& \ \bar \psi'(\tau) 
= e^{i \alpha (\tau)} \bar \psi (\tau) \cr
a (\tau) \ &\to& \ a'(\tau) = a (\tau) +\dot \alpha (\tau)
\eea
where the gauge transformations $e^{-i \alpha (\tau)} $
are required to be periodic functions on $[0,1]$.
Note also that in one dimension the only gauge 
invariant quantity that can be 
constructed from the gauge field is the Wilson loop 
\be
w= e^{i\int_0^1 d\tau\, a(\tau)}  \ .
\ee
Using ``small'' gauge transformations, i.e. those continuously connected 
to the identity, one can bring $a(\tau)$ to a constant value $\phi$
\be
\phi = \int_0^1 d\tau \, a(\tau) \ .
\ee
Then ``large'' gauge transformations with $\alpha(\tau) = 2 \pi n \tau$
allow to identify 
\be
\phi \sim \phi + 2 \pi n \ , \quad \quad n\ {\rm integer}.
\ee
Therefore one can take $\phi\in [0,2\pi]$ as the fundamental region 
of the moduli space.
The value of the Wilson loop is given by the phase
$w= e^{i \phi}$, and once again one can recognize that $\phi$ is an angle.

Let us now comment on the choice of antiperiodic boundary conditions 
for the fermions.
In the gauge $a(\tau)=\phi$ the action (\ref{act-fer}) becomes
\be
S \sim \int_0^1 d\tau\, \bar \psi (\dot \psi + i \phi  \psi) \ .
\label{3.7}
\ee
One may now redefine the fermion by
$\psi' = e^{i \phi \tau }\psi$
to eliminate the gauge field from the action
\be
S \sim \int_0^1 d\tau\, \bar \psi' \dot {\psi'} \ . 
\label{3.8}
\ee
However, the new field acquires twisted boundary conditions
\be
\psi' (1) = - e^{i \phi} \psi'(0)
\ee
so that the modulus $\phi\in [0,2\pi]$ interpolates between 
all possible boundary conditions specified by a phase.
Therefore there was no loss of generality in the original assumption
of antiperiodic boundary conditions:
the sum over spin structures is automatically taken care of
by the integration over the $U(1)$ modulus.
Note that a similar situation appears in the $N=2$ string theory
\cite{Mathur:1987uk,Ooguri:1991fp}.
For $\phi=\pi$ one obtains periodic boundary conditions. 
This is a delicate point, as the fermions acquire zero modes
(and the gravitinos develop corresponding moduli)
whose effects we will comment upon in the next section.

We are now ready to describe the gauge fixing of (\ref{25}).
We choose the gauge $\hat G=(\beta, 0,0,\phi)$, insert
the Faddeev-Popov determinants to eliminate 
the volume of the gauge group, and integrate over the moduli.
This gives  
\bea
\Gamma^{QFT}_{p}[g_{\mu\nu}] \eqa -{1\over 2}
\int_0^\infty {d \beta \over \beta}  \int_0^{2 \pi} {d \phi \over 2\pi}\,
\Big (2 \cos{\phi\over 2}\Big )^{-2} 
\int_{T^1} {\cal D} X\,  e^{-S[X, \hat G;g_{\mu\nu}]}
\label{34}
\eea
where:
$i)$ the measure over the proper time $\beta$ takes into account the effect 
of the symmetry generated by the Killing vector on the torus 
(namely the constant vector); 
$ii)$ the Faddeev-Popov determinants from the commuting susy ghosts are 
obtained from (\ref{24}) and are computed to give
$\det^{-1} (\partial_\tau +i\phi)\det^{-1} (\partial_\tau -i\phi)
=(2 \cos{\phi\over 2})^{-2} $. These are the inverse of the fermionic 
determinant arising from (\ref{3.7}) which is easily computed:  
antiperiodic boundary conditions produce a trace over the corresponding 
two-dimensional Hilbert space and thus 
$\det (\partial_\tau +i\phi) = e^{-i {\phi\over 2}}+e^{i {\phi\over 2}}=
2 \cos{\phi\over 2}$. For more details see the appendix.
$iii)$ The other Faddeev-Popov determinants 
do not give rise to any moduli dependent term.
$iv)$ The overall normalization $-1/2$ has been inserted to match QFT 
results. Up to the overall sign, one could argue that this factor 
is due to the fact that one is considering a real field rather 
than a complex one.  

Thus, up to the final integration over the moduli, one is left 
with a standard path integral for a nonlinear $N=2$ susy sigma model.
This path integral cannot be evaluated exactly for arbitrary background 
metrics $g_{\mu\nu}$, but it is the starting point 
of various approximations schemes.
In particular, we will consider here an expansion in terms of the proper
time $\beta$ which leads to the local heat-kernel expansion
of the effective action \cite{DW,DeWitt:2003pm}.
It is a derivative expansion depending 
on the so-called Seeley-DeWitt coefficients. Note that, 
strictly speaking, the effective action does not have 
a derivative expansion for massless fields, but the corresponding 
Seeley-DeWitt coefficients still characterize the field theoretical model.

\section{Proper time expansion}

In the previous section we have set up the worldline path integral
representation for the one-loop effective action of a $p$-form gauge field 
coupled to gravity.
We now wish to compute it in a proper time expansion.
For this purpose we need to evaluate, perturbatively
in $\beta $, the following path integral
\bea
\int_{T^1} {\cal D} X\,  e^{-S[X, \hat G;g_{\mu\nu}] +iq \phi}
\eea
where, for convenience,
we have extracted the constant Chern-Simons term from the action, so that
the nonlinear sigma model action reads as
\bea
S[X, \hat G;g_{\mu\nu}]
\eqa {1\over \beta} \int_0^1 d\tau \, 
\Big [ {1\over 2} g_{\mu\nu}(x) \dot x^\mu \dot x^\nu 
+   \bpsi_a (\dot \psi^a +  i \phi \psi^a +
\dot x^\mu \omega_\mu{}^a{}_b \psi^b) \ccr
&-& {1\over 2} R_{abcd} \bar\psi^a \psi^b \bar \psi^c \psi^d \Big ] \ .
\eea
We have found it convenient to scale the fermion by
$\psi^a \to \psi^a/ \sqrt{\beta}$ to extract a global factor ${1/\beta}$. 
This shows that $\beta$ can be used as a loop counting parameter and 
thus organizes the loop expansion.
The perturbative calculation now is standard and mimics
the worldline treatment of spin 0 and 1/2 particles described in
\cite{Bastianelli:2002fv,Bastianelli:2002qw,Bastianelli:2003bg}.
One can extract the dependence on the zero modes $x_0^\mu$
of the coordinates and obtains
\bea
\Gamma^{QFT}_{p}[g_{\mu\nu}] \eqa -{1\over 2}
\int_0^\infty {d \beta \over \beta}
\int_0^{2 \pi} {d \phi \over 2\pi}\,
\Big (2 \cos{\phi\over 2}\Big)^{D-2} e^{iq\phi}  
\int {d^D x_0 \sqrt{g (x_0)} \over (2 \pi \beta)^{D\over 2}}\, 
\big \la e^{-S_{int}} \big \ra \ .
\label{4.3}
\eea
Let us comment on the various terms appearing in this formula.\\ 
$i)$ 
The constant zero modes $x_0^\mu$ are factorized by setting 
$x^\mu(\tau)=x^\mu_0+ y^\mu(\tau)$ and imposing the Dirichlet boundary 
conditions $y^\mu(0)=y^\mu(1)=0$ on the quantum fields $y^\mu(\tau)$.
This describes a loop with a fixed point $x_0$. 

\[
{\raisebox{-1cm}
{\includegraphics*[266pt,654pt][348pt,732pt]{graphs.epsi}}}
\hspace{-2.615cm}
{\raisebox{-.14cm}
{\includegraphics*[205pt,680pt][234pt,695pt]{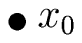}}}
\]
\begin{center}
{\bf Figure 3:} Loop with a marked point
\end{center}

\noindent 
The measure for integrating over $x_0$ is fixed by covariance and by 
checking the correct
flat space limit, so that $\la 1 \ra=1$  as far as the 
path integral over the $y$'s is concerned.
Other options for factorizing the zero modes are available
and have been carefully analyzed in \cite{Bastianelli:2003bg}.
\\  
$ii)$ 
The extra factor $(2 \cos{\phi\over 2})^{D}$
comes form the normalization of the fermionic path integral
and corresponds to $\det^{D} (\partial_\tau +i\phi)$.\\
$iii)$  
One can deduce the quantization law for the 
Chern-Simons coupling from (\ref{4.3}). 
The point $\phi=0$ is gauge equivalent to the point $\phi=2 \pi$.
Thus in even dimensions periodicity in $\phi$
requires that $q$ be an integer.
On the other hand, 
in odd dimensions it must be a half integer to compensate  
the anomalous behavior of the fermionic determinants
$\det^{D} (\partial_\tau +i\phi) =(2 \cos{\phi\over 2})^{D}$.\\
$iv)$  
The propagators are identified from the quadratic part of the action which 
is obtained by Taylor expanding the metric around the point $x_0^\mu$ 
\bea
S_{2}=
{1\over \beta} \int_0^1 d\tau \, 
\Big [ {1\over 2} g_{\mu\nu}(x_0) \dot y^\mu \dot y^\nu 
+ \bpsi_a (\dot \psi^a +  i \phi \psi^a ) \Big ]  \ .
\label{4.4}
\eea
The interactions are then given by $S_{int}= S - S_{2}$. 
In particular, the fermion propagator is 
\be
\langle \psi^a(\tau) \bar \psi_b(\sigma)\rangle =
\beta\, \delta^a_b \Delta_{AF}(\tau-\sigma,\phi) 
\ee
where the function $\Delta_{AF}(x,\phi)$ is given for $x\in ]-1,1[$ by 
\bea 
\Delta_{AF}(x,\phi)
= {e^{-i\phi x} \over 2 \cos {\phi\over 2}}
\Big [ e^{i{\phi\over 2}}\theta(x)  - e^{-i{\phi\over 2}}\theta(-x)\Big ] 
\eea
with $\theta(x)$ the step function 
(taking the value 0 for $x<0$ and 1 for $x>0$).
It satisfies
\bea
(\partial_x + i\phi) \Delta_{AF}(x,\phi) =\delta(x) \ .
\eea
For $\phi =0$  it reduces to the propagator 
for antiperiodic fermions already used in \cite{Bastianelli:2002qw},
$\Delta_{AF}(x) = {1\over 2}[\theta(x)  -\theta(-x)]$.
For coinciding points ($\tau =\sigma$ i.e. $x=0$) it 
takes the (regulated) value
\be
\Delta_{AF}(0,\phi) = {i\over 2} \tan {\phi \over 2}\ .
\ee
In the appendix we give some details on how this propagator is obtained.
\\
$v)$ It is convenient to  use ``measure'' ghosts to exponentiate 
the nontrivial dependence of the path integral measure on $g_{\mu\nu}$. 
These ghosts can be added by replacing 
\be 
\dot x^\mu \dot x^\nu \quad \to \quad \dot x^\mu \dot x^\nu + a^\mu a^\nu +
b^\mu c^\nu 
\label{4.9}
\ee
in the action.
The $a^\mu$ ghosts are commuting while the $b^\mu,c^\nu$ ghosts are 
anticommuting, and they keep track in all Feynman diagrams
of the contributions coming from the path integral measure associated 
to nonlinear sigma models \cite{Bastianelli:1991be,Bastianelli:1992ct}.
They make the calculation finite, but a regularization is still needed 
to remove finite ambiguities.
The relevant propagators are reported in the appendix.\\
$vi)$ 
To compute $\big \la e^{-S_{int}} \big \ra $
one must select a regularization scheme and add the corresponding 
counterterm. 
We employ dimensional regularization. 
We have checked that for our $N=2$ model the counterterm from 
dimensional regularization vanish, just like in the 
$N=1$ model \cite{Bastianelli:2002qw}.
Note that other regularization schemes may need a nonvanishing counterterm
(as the time-slicing scheme \cite{deBoer:1995cb}).
We discuss this issue more extensively in the appendix.
Finally, we choose  Riemann normal coordinates 
and a two-loop calculation  on the worldline  produces 
(we use conventions with $R>0 $ on the sphere and rewrite 
$\tan^2 \frac{\phi}{2}=\cos^{-2}\frac{\phi}{2} -1$ for convenience 
whenever necessary)
\bea
\big \la e^{-S_{int}} \big \ra 
\eqa 
1 + \beta R \Big(  \frac{1}{12}-\frac{1}{8}
\cos^{-2}\frac{\phi}{2} 
\Big )  
\nonumber \\ 
&+& \beta^2 \bigg{[} R^2_{abcd} \Big{(} \frac{1}{720}-
\frac{1}{192} \cos^{-2}\frac{\phi}{2} 
+ \frac{1}{128} \cos^{-4}\frac{\phi}{2} 
\Big{)} \nonumber \\  
&+& R^2_{ab} \Big{(} -\frac{1}{720}+\frac{1}{32} \cos^{-2}\frac{\phi}{2} 
- \frac{1}{32}\cos^{-4}\frac{\phi}{2} \Big{)} \nonumber \\  
&+& R^2 \Big{(}\frac{1}{288}-\frac{1}{96} \cos^{-2}\frac{\phi}{2} 
+  \frac{1}{128}\cos^{-4}\frac{\phi}{2} \Big{)} \nonumber \\ 
&+& \nabla^2 R \Big{(} \frac{1}{120}-\frac{1}{96}
\cos^{-2}\frac{\phi}{2} \Big{)} 
\bigg{]} + O(\beta^3)   \ .
\label{4.10}
\eea

We are now left to insert this perturbative result into eq. (\ref{4.3}). 
As already mentioned, infrared divergences prevent a meaningful
local expansion of the effective action for massless fields.
This is signaled from the fact  that when (\ref{4.10}) is inserted into 
(\ref{4.3}) the proper time integral does not converge at 
$\beta =\infty$ (the infrared region). In fact, the standard mass term 
$e^{- {1\over 2} m^2 \beta }$ which insures convergence for massive theories
is absent in this case. One may nevertheless write 
eq. (\ref{4.3}) in the form
\bea
\Gamma^{QFT}_{p}[g_{\mu\nu}] 
\eqa -{1\over 2}
\int_0^\infty {d \beta \over \beta} Z_p(\beta) \ccr
Z_p(\beta) \eqa
\int {d^D x_0 \sqrt{g (x_0)} \over (2 \pi \beta)^{D\over 2}}\, 
\Big ( a_0 + a_1 \beta +a_2\beta^2 + \ldots \Big )
\label{SDW}
\eea
where the coefficients $a_i$ are the Seeley-DeWitt coefficients
(in the coincidence limit).
Even if convergence of the proper time integral in the upper limit 
is not guaranteed, one can still compute these coefficients.
They characterize the theory. For example they identify the counterterms
needed to renormalize the full effective action.   
In addition, $a_2$ gives the trace anomaly for a spin 1 field 
in four dimensions, and $a_1$ the trace anomaly for a scalar field in  
two dimensions.
To compute these coefficients and test the correctness of the previous 
set up we must integrate over the $U(1)$ modulus.
Before doing that let us parametrize the structure of the Seeley-DeWitt 
coefficients contained in the round bracket of $Z(\beta)$ as follows   
\be
\Big ( v_1 + v_2 R \beta 
+ (v_3 R^2_{abcd}+v_4 R_{ab}^2+ v_5 R^2 + 
v_6 \nabla^2 R)\beta^2  + O(\beta^3) \Big ) \ .
\ee
Next we will compute the coefficients $v_i$.

\subsection{Seeley-DeWitt coefficients} 
\label{sec:4.1}

Let us now discuss the integration over the modular parameter $\phi$.
We see that at $\phi=\pi$ there is a potential divergence appearing
in the integrand (\ref{4.10}) (as $\cos^{-2} {\pi\over 2}=\infty$).   
In fact at this point of moduli space the fermions 
develop a zero mode and, as a consequence, 
their propagator develops a singularity.
According to our previous discussion, the point $\phi=\pi$ 
corresponds to periodic boundary conditions and 
the kinetic operator in (\ref{3.8}) can only be inverted
in the space orthogonal to the space of constant fields.
Note that at this particular point of moduli space 
the gravitinos cannot be completely gauged away 
and fermionic moduli appears (the constant gravitinos).

To take into account this singular point  we are going to use an
analytic regularization in moduli space.
First we change coordinates and use the Wilson loop 
variable $w=e^{i\phi}$ instead of $\phi$.
The integration region of this new variable is the unit circle
$\gamma$ on the complex $w$-plane
\be
\int_0^{2 \pi} {d\phi\over 2\pi} = \oint_\gamma {dw \over 2 \pi i w} \ .
\ee
The singular point $\phi =\pi$ is now mapped to $w=-1$. 
Our prescription is to use complex contour integration and deform the 
contour to exclude the point $w=-1$ (say by moving it outside 
to $w=-1-\epsilon$ with $\epsilon >0$, 
and then letting $\epsilon \to 0^+$). 

\[
\raisebox{-1cm}{\scalebox{.5}{
{\includegraphics*[1pt,1pt][290pt,290pt]{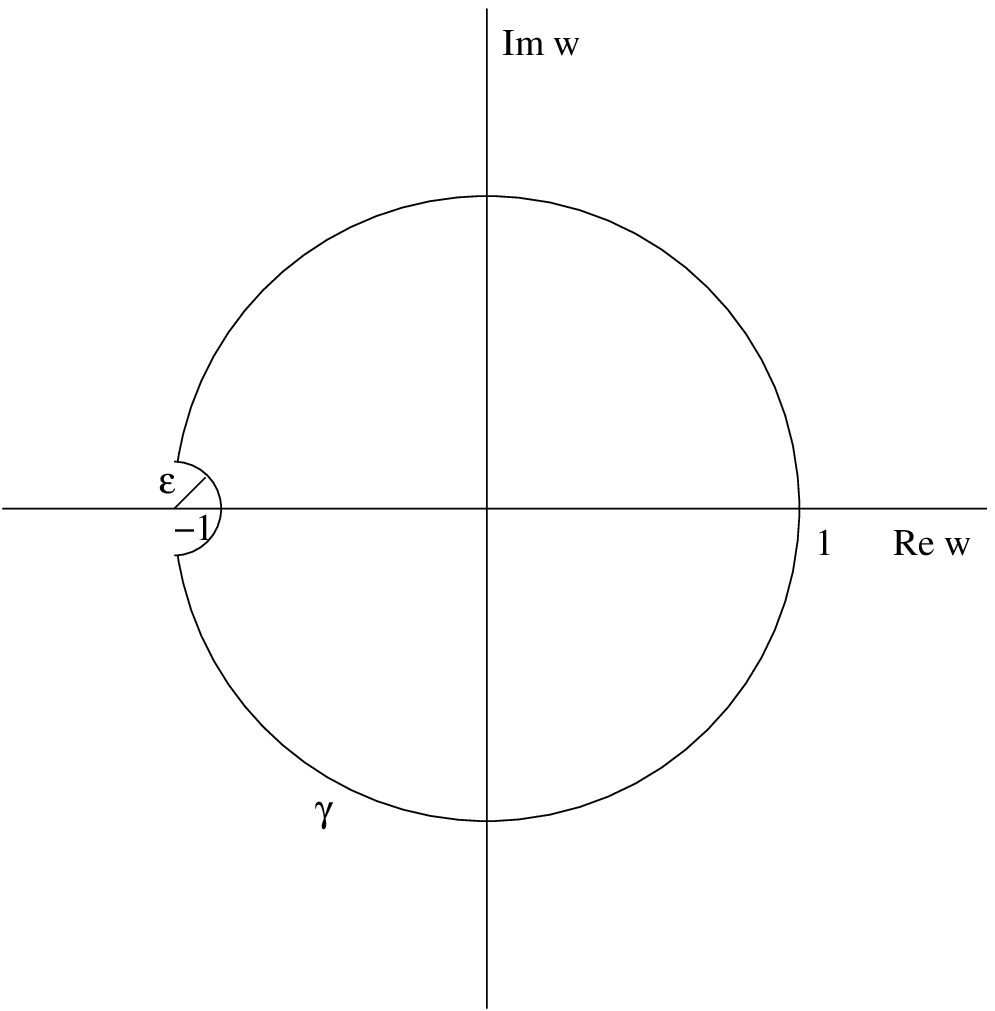}}}}
\]
\begin{center}
{\bf Figure 4:} Regulated contour for the $U(1)$ modular parameter
\end{center}

\noindent
This prescription allows us to recover all known results, though
a deeper justification would be welcome.
We may note that this prescription 
can be interpreted as giving the worldline fermions a small mass $\epsilon$
to lift the zero modes appearing at the point $\phi=\pi$ of moduli
space, i.e. replacing  $\phi=\pi$ by $\phi=\pi-i\epsilon$.
However, it does not seem obvious to us why one should require $\epsilon >0$.

The computation proceeds now as follows. 
From
\bea 
Z_p(\beta)&=&\int \frac{d^D x \sqrt{g}}{(2 \pi \beta)^{\frac{D}{2}}} 
\int_0^{2\pi} \frac{d\phi}{2\pi }\,
e^{i q \phi } \Big(2\cos \frac{\phi}{2} \Big )^{D-2} 
\big \la e^{-S_{int}} \big \ra 
\label{4.14}
\eea 
with the expression (\ref{4.10}) 
one can get all coefficients $v_1, ..., v_6$ for 
arbitrary $(D,p)$. They are combinations of the following basic 
integrals (recall that $q=\frac{D}{2} -p-1$)
\bea 
I_n(D,p) &\equiv& 2^{D-2} 
\int_0^{2\pi} \frac{d\phi}{2\pi }\, 
e^{i q \phi } \Big (\cos \frac{\phi}{2} \Big )^{D-2n} 
\nonumber \\[2mm] 
  &=& 2^{2n-2} \oint_{\gamma} \frac{dw}{2 \pi i} 
\frac{(1+w)^{D-2n}}{w^{p+2-n}}   \ .
\label{4.15}
\eea 
As already mentioned, there 
is a possible pole at $w=-1$. With the prescription to push the 
pole out of the integration circle, the integrals are given  
by computing the residue at the pole $w=0$, and the result is 
\bea 
&&I_n(D,p) = \frac{2^{2n-2}}{(p+1-n)!} \frac{d^{p+1-n}}{dw^{p+1-n}} 
(1+w)^{D-2n}\Big |_{w=0}  \quad \; \mbox{if} \quad p \ge n-1 
\ccr[2mm]
&&I_n(D,p) = 0  \qquad \qquad \qquad \qquad \qquad \qquad 
\qquad  \quad  
\qquad\mbox{if} \quad p < n-1   \ .
\label{4.17}
\eea 
For the coefficients $v_i$ one then gets
\bea 
&& 
v_1(D,p)=I_1(D,p)= \frac{(D-2)!}{p!(D-2-p)!} 
\ccr[2mm] 
&& 
v_2(D,p) = \frac{1}{12}I_1(D,p) -\frac{1}{8}I_2(D,p) 
\ccr[2mm]  
&& 
v_3(D,p) = \frac{1}{720}I_1(D,p) -\frac{1}{192}I_2(D,p)+\frac{1}{128}I_3(D,p) 
\ccr[2mm]  
&& 
v_4(D,p) = -\frac{1}{720}I_1(D,p) +\frac{1}{32}I_2(D,p)-\frac{1}{32}I_3(D,p) 
\ccr[2mm] 
&& 
v_5(D,p) = \frac{1}{288}I_1(D,p) -\frac{1}{96}I_2(D,p)+\frac{1}{128}I_3(D,p) 
\ccr[2mm]  
&& 
v_6(D,p) = \frac{1}{120}I_1(D,p) -\frac{1}{96}I_2(D,p)  \ .
\label{4.18}
\eea 
Using (\ref{4.17}) one may check that all known values for 
$p=0,1,2$ are reproduced, see for example  
\cite{DW} and \cite{DeWitt:2003pm}.
In particular, it is worth noticing the case of $p=2$, which was derived
after considerable algebra in volume II of \cite{DeWitt:2003pm}
(see page 974).
For completeness and future reference let us list these coefficients
in the format
\be
F_{p+1}  \to  (v_1; v_2; v_3; v_4; v_5; v_6) \ . 
\ee
For $p=-1,0,1,2,3$ we have
\bea
F_0 &\to & 
\Big (0 ; 0 ; 0 ; 0 ; 0 ; 0 \Big ) \ccr[3mm]
F_1 &\to & 
\Big (1 ; 
{1\over 12} ; 
{1\over 720} ; 
-{1\over 720} ; 
{1\over 288} ; 
{ 1\over 120} \Big ) \ccr[3mm]
F_2 &\to & 
\Big (D-2; 
{D-8\over 12};
{D-17\over 720}; 
{92-D\over 720}; 
{D-14\over 288};
{ D-7\over 120} \Big ) \ccr[3mm]
F_3  &\to & 
\Big ({(D-2)(D-3)\over 2}; 
{D^2 -17 D + 54 \over 24}; 
{(D-17)(D-18) \over 1440}; 
\ccr
&&   
-{D^2 -185 D + 1446\over 1440};
{D^2 -29 D + 174\over 576}; 
{D^2 -15 D + 46\over 240} \Big ) \ccr[3mm]
F_4  
&\to & 
\Big (
{1\over 6}(D-2)(D-3)(D-4);
{1\over 72}(D-4)(D^2 -23 D + 96);
\ccr
&&   
{1\over 4320}(D^3 -54 D^2 + 971 D -4164);
-{1\over 4320}(D^3 -279 D^2 + 4616 D -18384);
\ccr
&& {1\over 1728}(D-12)(D^2 -33 D + 170);
{1 \over 720} (D-4)(D^2 -20 D + 81)
\Big) \ .
\label{4.20}
\eea
The case $F_1$ corresponds to a massless scalar 
(without any improvement term), and the case $F_2$
to a massless spin 1 field. The latter contains, in particular, the correct
trace anomaly of the photon in $D=4$, already obtained with a worldline 
approach in \cite{Bastianelli:1992ct}, but with a different quantum 
mechanical model (which we shall consider in subsection \ref{sec:4.3}). 
The case $F_3$ describes a 2-form gauge potential which 
is the potential that couples
naturally to a string source. It is conformally invariant in $D=6$, 
and its trace anomaly has been computed in \cite{Bastianelli:2000hi}. 
The latter is encoded in the coefficient $a_3$,
and could be obtained with worldline methods by computing
the next perturbative correction to (\ref{SDW}).
The case $F_4$ seems to be a new result, as we have not been able to find it 
in the literature.
The coefficient $v_1$ counts the number of physical  degrees 
of freedom and it is easily checked to be correct for any $(D,p)$.

\subsection{An application: $F_5$ and $F_6$} 
\label{sec:4.2}

As an application of the previous general result
we can make explicit the coefficients for the 4- and 
5-form, which, to our knowledge, are not present in the literature.  
 For the $p=4$ case, we need the following values 
\bea 
I_1(D,4) &=& {(D-2)(D-3)(D-4)(D-5) \over 24} \\  
I_2(D,4) &=& 4 {(D-4)(D-5)(D-6) \over 6} \\ 
I_3(D,4) &=& 16 {(D-6)(D-7) \over 2}   \ .
\eea 
Using the formulas given above we get for $F_5=dA_4$ 
\bea 
F_5   
&\to &  
\Big ( 
{1\over 24}(D-2)(D-3)(D-4)(D-5);
\ccr
&&
{1\over 288}(D-4)(D-5)(D^2 -29 D + 150); 
\ccr 
&&    
{1\over 17280}(D^4 -74 D^3 + 2051 D^2 -18634D + 52680);
\ccr 
&& 
-{1\over 17280}(D^4 -374 D^3 + 9791 D^2 -82954D + 224760);
\ccr 
&&   {1\over 6912}(D^4 -62 D^3 + 1223 D^2 - 9322 D +24024); 
\ccr 
&& 
{1 \over 2880} (D-4)(D-5)(D-7)(D-18) 
\Big)   \ .
\eea 
To test these coefficients, one may check that $F_5 \sim F_1$ in $D=6$,
as in such dimensions a 4-form gauge field gives a dual description of 
a scalar field. 
In $D=6$ the topological Euler density appears  
at higher order in $\beta$ so the duality holds exactly
(in $D=4$ a 2-form gives a dual description of a scalar field 
only up to the topological Euler density \cite{Duff:1980qv},
the appearance of the Euler term is a general phenomenon
which we discuss in section \ref{sec:5}).

Moving on to the $p=5$ case, we need the following integrals 
\bea 
I_1(D,5) &=& {(D-2)(D-3)(D-4)(D-5)(D-6) \over 120} \\  
I_2(D,5) &=& 4 {(D-4)(D-5)(D-6)(D-7) \over 24} \\ 
I_3(D,5) &=& 16 {(D-6)(D-7)(D-8) \over 6}  
\eea 
and the explicit coefficients for $F_6=dA_5$ turn out to be 
\bea 
F_6   
&\to &  
\Big ( 
{1\over 120}(D-2)(D-3)(D-4)(D-5)(D-6);
\ccr 
&& 
{1\over 1440}(D-4)(D-5)(D-6)(D-8)(D-27);
\ccr 
&&    
{1\over 86400}(D-6)(D^4 -89 D^3 + 3071 D^2 -33379D + 111420);
\ccr 
&& 
-{1\over 86400}(D-6)(D^4 -464 D^3 + 14471 D^2 -145504 D + 466320); 
\ccr 
&&   {1\over 34560}(D-6)(D^4 -74 D^3 + 1751 D^2 - 15934 D +48840); 
\ccr 
&& 
{1 \over 14400} (D-4)(D-5)(D-6)(D^2 -30 D + 181) 
\Big) \ .   
\eea 
These coefficients pass the expected duality tests. 
An immediate check to perform is to see that $F_6 \sim F_0\sim 0$ in $D=6$, 
as $F_6$ in six dimensions is dual to a scalar field strength
which has no gauge potential. 
As a further check, one can for example verify 
that $F_6 \sim F_2$ in $D=8$. 
Again, in $D=6$ and $D=8$ the expected dualities hold exactly since 
the topological Euler density appears at higher order
in the proper time expansion.

\subsection{Another application: susy ungauged, $A_4$ and $A_5$} 
\label{sec:4.3}

The Seeley-DeWitt coefficients for the  differential $(p+1)$-form $A_{p+1}$, 
not interpreted as a field strength, but as a differential form with 
kinetic operator 
given by the generalized laplacian $dd^\dagger + d^\dagger d$,
can be obtained from the $N=2$ model with ungauged susy 
(gauging susy enforces Maxwell equations and Bianchi identities). 
This corresponds to changing the factor 
$\big (2\cos\frac{\phi}{2} \big)^{D-2} \to \big(2\cos \frac{\phi}{2} \big)^D$
in (\ref{4.14}). The coefficients  
will still be combinations of the basic integrals given in eq. 
(\ref{4.15}) where we only need to replace $2^{D-2} \to 2^D$
as overall normalization factor.
In this case $I_0$ will also enter the computation
together with $I_1$ and  $I_2$. 
In this way, one can reproduce for $p=-1,0,1$ (remember that 
the Chern-Simons coupling  $q={D\over 2}-p-1$ selects a $(p+1)$-form) 
the coefficients for $A_0,A_1,A_2$  already present 
in the literature, see again \cite{DW} and \cite{DeWitt:2003pm}.
We list them here for completeness
\bea 
A_0 &\to & 
\Big (1; {1\over 12} ; {1\over 720} ; -{1\over 720} ; 
{1\over 288} ; { 1\over 120} \Big ) \ccr[3mm]
A_1 &\to & 
\Big (D ; {D-6\over 12} ; {D-15\over 720} ; {90-D\over 720} ; 
{D-12\over 288} ; { D-5\over 120} \Big ) \ccr[3mm]
A_2 &\to & 
\Big ({D(D-1)\over 2} ; 
{D^2 -13 D + 24 \over 24} ; 
{D^2 -31 D + 240 \over 1440} ; 
\ccr
&&   
-{D^2 -181 D + 1080\over 1440} ;
{D^2 -25 D + 120\over 576} ; 
{D^2 -11 D + 20\over 240} \Big ) \ .
\eea
However, the $N=2$ model has produced the result for all differential forms. 
As an example, we can make explicit the cases with $p=2,3,4$ to obtain the 
coefficients for $A_3$, $A_4$ 
and $A_5$ which are not given in the literature. 
We get    
\bea
A_3 &\to & 
\Big ( {1\over 6} D(D-1)(D-2) ;
{1\over 72} (D-2)(D^2-19D+54) ;
\ccr
&&
{1\over 4320} (D^3 -48 D^2+767 D-2430) ;
\ccr
&&
-{1\over 4320} (D^3 -273 D^2+3512 D-10260) ;
\ccr
&& 
{1\over 1728} (D-10)(D^2-29 D+108) ;
\ccr
&&
{1\over 720} (D-2)(D^2-16 D+45)\Big) 
\eea
\bea 
A_4   
&\to &  
\Big ( 
{1\over 24}D(D-1)(D-2)(D-3) ;
\ccr 
&& 
{1\over 288}(D-2)(D-3)(D^2 -25 D + 96) ;
\ccr 
&&    
{1\over 17280}(D^4 -66 D^3 + 1631 D^2 -11286 D + 23040) ; 
\ccr 
&& 
-{1\over 17280}(D^4 -366 D^3 + 7571 D^2 -48246 D + 95040) ; 
\ccr 
&&   {1\over 6912}(D^4 -54 D^3 + 875 D^2 - 5142 D + 9792) ;
\ccr 
&& 
{1 \over 2880} (D-2)(D-3)(D-5)(D-16)
\Big)   
\eea 
\bea 
A_5   
&\to &  
\Big ( 
{1\over 120}D(D-1)(D-2)(D-3)(D-4) ;
\ccr 
&& 
{1\over 1440}(D-2)(D-3)(D-4)(D-6)(D-25) ; 
\ccr 
&&    
{1\over 86400}(D-4)(D^4 -81 D^3 + 2561 D^2 -22131 D + 56250) ;
\ccr 
&& 
-{1\over 86400}(D-4)(D^4 -456 D^3 + 11711 D^2 -93156 D + 229500) ;
\ccr 
&&   {1\over 34560}(D-4)(D^4 -66 D^3 + 1331 D^2 - 9786 D + 23400) ; 
\ccr 
&& 
{1 \over 14400} (D-2)(D-3)(D-4)(D^2 -26 D + 125) 
\Big)   \ .
\eea 
As a test, one may check that Poincar\'e duality holds, for example 
$A_4 \sim A_2$ and $A_5 \sim A_1$ in $D=6$.

As a final test of our results, one may note that these differential forms 
with generalized laplacian as kinetic operator appear in the covariantly 
gauge fixed action for an antisymmetric tensor gauge field.
Denoting by  $W_p$ the one-loop effective action of one such 
a $p$-form and recalling the ``triangular'' structure of the gauge fixed
action \cite{Siegel:1980jj,Thierry-Mieg:1980it}
one can identify
\bea
\Gamma^{QFT}_{p} \eqa W_{p} -2 W_{p-1}+ 3 W_{p-2}+ \ldots 
+ (-1)^p (p+1) W_0
\ccr
\eqa \sum_{k=0}^p  (-1)^k (k+1)W_{p-k} \ .
\label{4.32}
\eea
To check this relation, let us remember that the Seeley-DeWitt 
coefficients which characterize $W_{p}$
can be obtained by taking (\ref{4.18}), which gives the coefficients for 
$\Gamma^{QFT}_{p}$, and  replacing everywhere 
$I_n(D,p) \rightarrow 4I_{n-1}(D,p-1)$. 
Having noticed this, eq. (\ref{4.32}) can be seen as a consequence of the 
following identity involving the integrals in (\ref{4.15})
\bea
I_n(D,p) &=& {2^{2n-2} \over (p+1-n)!} \partial_w^{p+1-n} 
(1+w)^{D-2n}\Big |_{w=0} \nonumber \\
     &=& {2^{2n-2} \over (p+1-n)!} \partial_w^{p+1-n} 
\bigg{[}(1+w)^{D-2n+2}(1+w)^{-2}\bigg{]}\Big |_{w=0} \nonumber \\
     &=& 2^{2n-2} \sum_{k=0}^{p+1-n} {\bigg{[} 
\partial_w^{p+1-n-k} (1+w)^{D-2n+2}\bigg{]}
\bigg{[} \partial_w^{k} (1+w)^{-2}\bigg{]} 
\over (p+1-n-k)!k!}\Big |_{w=0} \nonumber \\
  &=& \sum_{k=0}^{p+1-n} \bigg{[} 4I_{n-1}(D,p-k-1)\bigg{]} (-1)^k (k+1)  \ .
\eea
Using that $I_{n-1}(D,p-k-1)$ is zero if $k>p+1-n$, see (\ref{4.17}), 
we may equivalently write this identity as
\bea
I_n(D,p) = \sum_{k=0}^{p} (-1)^k (k+1)  4I_{n-1}(D,p-k-1) 
\eea
from which eq. (\ref{4.32}) follows.

\section{Discussion and conclusions}
\label{sec:5}

We have used the $N=2$ spinning particle to compute one-loop
effects due to the propagation of differential forms coupled to gravity,
including as a particular case a spin 1 field.
One of the most interesting points of the construction is the appearance 
of the $U(1)$ modulus $\phi$, a parameter which does not emerge in
the standard derivation of Feynman rules.
This is also a delicate point which deserves further discussions.
For this purpose it is convenient to switch 
to an operatorial picture and cast the effective action
(\ref{34}) in the form
\bea
\Gamma^{QFT}_{p}[g_{\mu\nu}] \eqa -{1\over 2}
\int_0^\infty {d \beta \over \beta}  \int_0^{2 \pi} {d \phi \over 2\pi}\,
\Big (2 \cos{\phi\over 2}\Big )^{-2} 
{\rm Tr}\, \big [e^{i\phi (\hat N - {D\over 2}+q)}
e^{-\beta  \hat H} \big ]
\eea
where the path integral on the torus has been represented by the 
trace in the matter sector of the Hilbert space of the spinning particle 
(i.e. excluding the ghost sector).
Here $\hat N = \hat \psi^\mu \hat \psi^\dagger_\mu $
is the (anti) fermion number operator (it counts the degree of the 
field strength form, and up to the ordering and to
the Chern-Simons coupling coincides with the current $-\hat J$)
and $\hat H$ is the quantum hamiltonian
without the coupling to the gauge field, which has been explicitly factorized.
Computing the trace and using the Wilson loop variable $w=e^{i\phi}$
gives an answer of the form
\bea
\Gamma^{QFT}_{p}[g_{\mu\nu}] \eqa -{1\over 2}
\int_0^\infty {d \beta \over \beta}  
\oint_{\gamma} \frac{dw}{2 \pi i w} 
\frac{w}{(1+w)^2} \sum_{n=0}^D  w^{n-{D\over 2} +q}\, t_n(\beta)
\label{5.2}
\eea
where the coefficients $t_n(\beta)$ arise from the trace  restricted 
to the sector of the Hilbert space with occupation number $n$.
From the answer written in this form one can make various comments.

If susy is not gauged then the ghost term $\frac{w}{(1+w)^2}$ is absent, 
and one recovers the model described in section \ref{sec:4.3}. Then, 
there is no pole at $w=-1$, at least for finite $D$,
and the $w$ integral projects onto the sector of 
the Hilbert space with occupation number $n-{D\over 2} +q =0$, i.e. $n=p+1$.
This describes a $(p+1)$-form with generalized laplacian as kinetic 
operator. The absence of the pole at $w=-1$ means that excluding or 
including this point in the regularized
contour $\gamma$ must produce the same answer.
This is related to  Poincar\'e duality, which tells that the result for 
a $(p+1)$-form must be equivalent to that of a  $(D-p-1)$-form.
The change from $(p+1)$ to  $(D-p-1)$ in (\ref{5.2}) is described by
$q \to -q$, which can be undone by a change of integration
variable  $w \to w'= {1\over w}$ (or $\phi\to -\phi$).
This proves Poincar\'e  equivalence, i.e. $t_n(\beta)=t_{D-n}(\beta)$.

Next consider the case of gauged susy. Now one must include the 
contribution of the ghosts' determinants $\frac{w}{(1+w)^2}$  
and face the appearance of a possible pole at $w=-1$.
The duality between a gauge $p$-form and a gauge $(D-p-2)$-form is again
described by $q \to -q$ and compensated by the 
change of integration variable $w \to w'= {1\over w}$.
However, the original contour which was regulated by excluding
the pole at $w=-1$ gets mapped into a contour which now
includes that pole, see figure 5. 

\[
\raisebox{-1cm}{\scalebox{.5}{
{\includegraphics*[1pt,1pt][290pt,290pt]{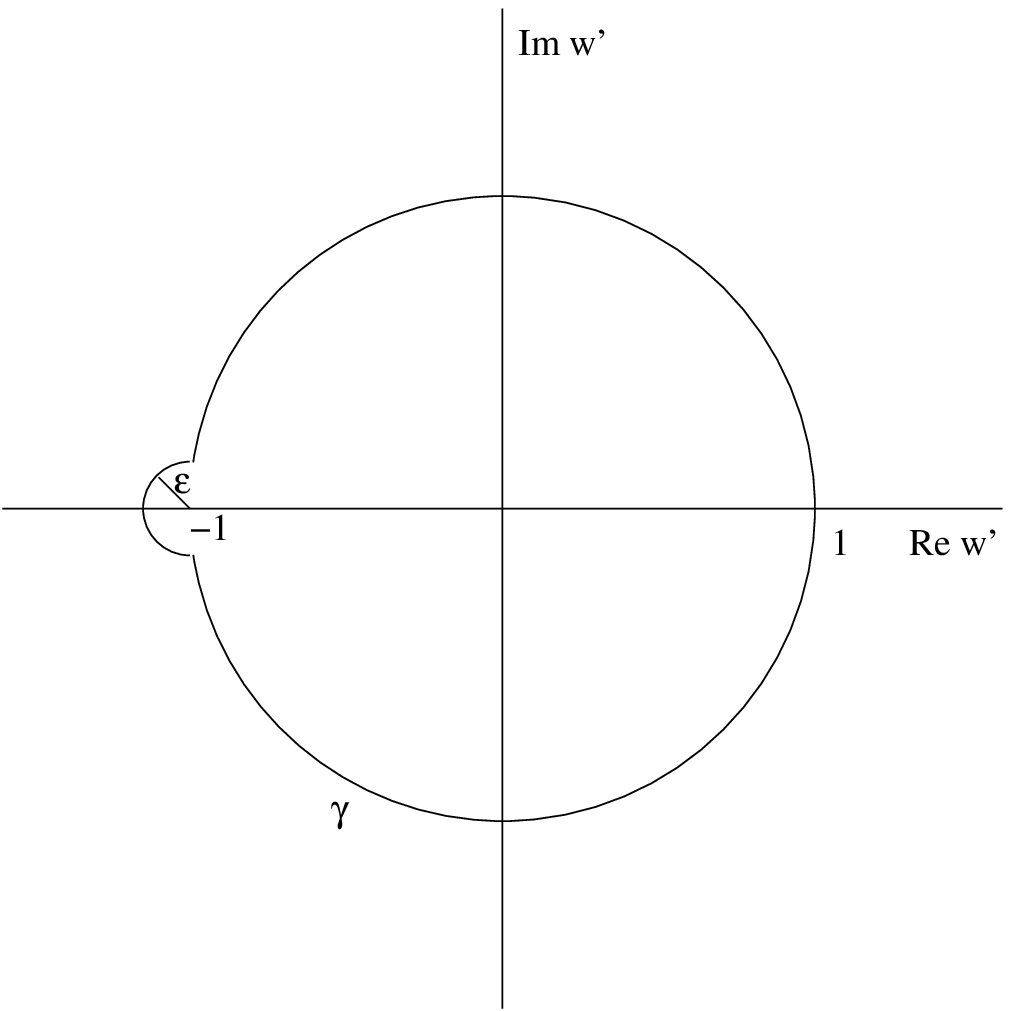}}}}
\]
\begin{center}
{\bf Figure 5:} Regulated contour in the variable $w'$
\end{center}

\noindent
Therefore strict equivalence is not guaranteed.
The mismatch corresponds to the residue at the pole $w=-1$, and 
can be computed as follows
\bea
&&
{\rm Res}\left ( \frac{1}{(1+w)^2} 
{\rm Tr}\, [ w^{\hat N -{D\over 2} +q }\,  e^{-\beta \hat H}], w=-1 
\right ) 
= {d\over d w} {\rm Tr}\, [ w^{\hat N -p -1 } e^{-\beta \hat H}]
\Big|_{w=-1} 
\ccr[1.8mm]
&& \quad \quad 
= {\rm Tr}\big [\big (\hat N -p-1\big ) 
\, (-1)^{\hat N -p- 2} \, e^{-\beta \hat H} \big ] 
\ccr[2mm]
&& \quad \quad 
=(-1)^{p}\, 
{\rm Tr}\, \big [ \hat N \, (-1)^{\hat N } \, e^{-\beta \hat H} \big ] 
+ (-1)^{p+1} (p+1)
{\rm Tr}\, \big [(-1)^{\hat N } \, e^{-\beta \hat H} \big ] 
\ccr[2mm]
&& \quad \quad =
(-1)^{p+1} Z_{D-1}(\beta) + (-1)^{p+1} (p+1) \chi \ .
\label{5.3}
\eea 
In fact, the second term in the last-but-one line is proportional 
to the Witten index \cite{Witten:1982df}, 
which is $\beta$ independent and computes the topological Euler number 
$\chi$ of target space  \cite{Alvarez-Gaume:1983at}.
The first term is instead
similar to an index introduced in  \cite{Cecotti:1992qh}
for two dimensional field theories, and which in our case
computes the partition function
$Z_{D-1}(\beta)$ of a $(D-1)$-form $A_{D-1}$
(with the top form $F_D$ as field strength)
\bea
&& {\rm Tr}\Big [\hat N  
\, (-1)^{\hat N} \, e^{-\beta \hat H} \Big ] 
= 
\sum_{n=1}^D (-1)^n\, n\, t_n(\beta)
=
\sum_{n=1}^D (-1)^n\, n\, t_{D-n}(\beta)
\ccr
&& \quad\quad =
 - \sum_{n=0}^{D-1} (-1)^n (n+1)\, t_{(D-1)-n}(\beta)  
= -Z_{D-1}(\beta)
\label{5.4}
\eea
where we have used Poincar\'e duality ($t_n=t_{D-n}$)
and recognized the ``triangular'' structure (\ref{4.32})
for the gauge field $A_{D-1}$.
The notation for the partition function
$Z_{D-1}(\beta)$ is as in (\ref{SDW}).
Thus we arrive at the following equivalence 
between propagating $p$-forms and $(D-p-2)$-forms 
\be
Z_{p}(\beta) = Z_{D-p-2}(\beta) 
+ (-1)^{p} Z_{D-1} (\beta) + (-1)^{p} (p+1)\chi  \ .
\label{5.5}
\ee
As the gauge field $A_{D-1}$ does not have propagating
degrees of freedom, the first Seeley-DeWitt coefficient $a_0$
is not spoiled by this duality.

For even $D$ the partition function $Z_{D-1} (\beta)$ 
is proportional to the Euler number, 
namely $Z_{D-1} (\beta) = -{D\over 2} \chi$,
as can be checked by using
Poincar\'e duality, so that equation (\ref{5.5})
simplifies to
\be
Z_{p}(\beta) = Z_{D-p-2}(\beta) 
+ (-1)^{p} \Big (p+1-{D\over 2}\Big )\chi  
\ee
or, equivalently,
\be
F_{p+1}\sim F_{D-(p+1)} +  (-1)^{p} \Big (1-{2(p+1)\over D} \Big )F_D  \ .
\label{5.6}
\ee
Thus we see that the mismatch is purely topological, 
as already noticed in \cite{Duff:1980qv}
for the duality between a scalar and  2-form gauge field in $D=4$.
The $\beta$ independence of $\chi$ shows, for example, 
that a mismatch between a scalar and a 4-form in $D=6$  
will be visible in the coefficient $a_3$, and so on.
The coefficient $a_3$ is laborious to compute, nevertheless
the bosonic ($N=0$) worldline coefficient has been already
calculated in \cite{Bastianelli:2000dw},
and could with some effort be dressed up to the $N=2$
model to obtain the $a_3$ coefficient for {\em all}
differential forms in {\em all} dimensions.
Note that (\ref{5.6})
is consistent with the selfduality of the $F_{D/2}$ form.

For odd $D$ the Euler number vanishes, and one has 
(at the level of field strengths) a duality of the form
\be
F_{p+1}\sim F_{D-(p+1)} +  (-1)^{p} F_{D}  \ .
\ee
One can easily check this relation in few examples, like
$F_1\sim F_4 + F_5$ and $F_2\sim F_3 - F_5$ in $D=5$. 
Another example is 
$F_1\sim F_2 + F_3$ in $D=3$, where one should remember
to use the $D=3$ identity relating the Riemann  
to the Ricci tensor which makes the four dimensional 
Gauss-Bonnet combination vanish identically,
$R_{abcd}^2 -4R_{ab}^2 + R^2=0$.
In ref. \cite{Schwarz:1984wk} it was noted that after integrating 
over $\beta$ to obtain the effective action, this mismatch can be 
related to the Ray-Singer torsion, which is also known to be a topological 
invariant.

Finally, one should mention that these inequivalences are present at
the level of unregulated effective actions.
They are given by local terms that can be subtracted in the renormalization
process, and thus, according to \cite{Siegel:1980ax},
they do not spoil the expected duality.

To conclude, we have seen that a worldline perspective on particles of 
spin 1 and on antisymmetric tensor gauge fields of arbitrary rank
coupled to gravity has produced a quite interesting 
and useful representation of the one-loop effective action.
We have tested that such a representation is correct by checking that 
the model correctly reproduces known 
Seeley-DeWitt coefficients, and in fact produces many more 
previously unknown ones.  
Other applications and extensions of this worldline approach
will be left for future work. 

\acknowledgments 

The research of FB has been sponsored in part by the 
Italian MIUR grant PRIN-2003023852 
``Physics of fundamental interactions: gauge theories, gravity and strings''
and by the EU network EUCLID (HPRN-CT-2002-00325).
The work of SG was partially supported by NSF grant no. PHY-0354776.
Finally, FB would like to thank the organizers of the
Simons Workshop on Mathematics and Physics 2004
at SUNY at Stony Brook for the stimulating environment and hospitality.

\appendix
\section{Appendix}
\label{section:appendix}

\subsection{Propagators and determinants}
\label{Propagators}

The relative coordinates $y^\mu(\tau)=x^\mu(\tau)-x^\mu_0$ and 
the measure ghosts $a^\mu, b^\mu, c^\mu$ are taken to vanish at the 
boundary of the space $[0,1]$. Then their propagators are
obtained from (\ref{4.4}), with measure ghosts inserted by
the shift (\ref{4.9}), and read
\bea
\la y^\mu (\tau) y^\nu(\sigma)\ra &=& 
- \beta g^{\mu\nu}(x_0)
\Delta(\tau,\sigma)
\nonumber \\ 
\la a^\mu(\tau) a^\nu(\sigma)\ra &=& 
 \beta g^{\mu\nu}(x_0) \Delta_{gh}
(\tau,\sigma) \label{propagdbc}
\nonumber \\ 
\la b^\mu(\tau) c^\nu (\sigma)\ra &=& 
-2 \beta  g^{\mu\nu}(x_0) \Delta_{gh}(\tau,\sigma)
\eea
with the functions $\Delta$ and $\Delta_{gh}$ given by
\bea
\Delta (\tau,\sigma) &=& 
 \sum_{m=1}^{\infty}
\biggl [ {- 2 \over {\pi^2 m^2}} \sin (\pi m \tau)
\sin (\pi m \sigma) \biggr ] =
(\tau-1)\sigma\, \theta(\tau-\sigma)+(\sigma-1)\tau\, \theta(\sigma-\tau)
\nonumber \\
\Delta_{gh}(\tau,\sigma) &=& 
\sum_{m=1}^{\infty}
2 \, \sin (\pi m \tau) \sin (\pi m \sigma) = 
\partial_\tau^2\Delta(\tau,\sigma) =
\delta(\tau,\sigma)
\eea
where $\theta(\tau-\sigma)$ is the standard step function and 
$\delta(\tau,\sigma)$ is the delta function which vanishes at the 
boundaries $\tau,\sigma =0,1$. 
These functions are not translationally invariant.
One could also use translational invariant Green functions
(the so-called ``string inspired'' propagators),
which are quite efficient in computations, but one has to 
remember to include an extra Faddeev-Popov determinant due to the
slightly more  complicated factorization
of the zero modes $x_0^\mu$ 
\cite{Friedan:1980jm,Kleinert:2003zq,Bastianelli:2003bg}.

The fermionic fields with antiperiodic boundary conditions 
can be expanded in half-integer modes
\be
\psi^a
(\tau) = \sum_{r\in Z+{1\over 2}} \psi^a_r \, {\rm e}^{2  \pi i r \tau} 
\ , \quad  \bar \psi^a (\tau) = 
\sum_{r\in Z+{1\over 2}} \bar \psi^a_r \, {\rm e}^{-2  \pi i r \tau} \ . 
\ee
Then from the action
\be
S= {1\over \beta} \int_0^1 d\tau \, 
\bpsi_a (\partial_\tau + i \phi) \psi^a  
\label{A.4}
\ee
one finds the propagator (AF stands for antiperiodic fermions)
\be
\la \psi^a(\tau) \bar \psi_b(\sigma)\ra 
= \beta \delta^a_b \Delta_{AF}(\tau-\sigma, \phi)\ ,
\ \ \ \ 
\Delta_{AF}(x,\phi) = \sum_{r\in Z+{1\over 2}} 
{-i\over 2 \pi  r + \phi }\,  e^{2  \pi i r x} 
\ee
which satisfies
\bea
(\partial_x + i\phi) \Delta_{AF}(x,\phi) 
= \sum_{r\in Z+{1\over 2}}  e^{2  \pi i r x} 
= \delta_{AF}(x) 
\eea
where $ \delta_{AF}$ is the delta function on the space of 
antiperiodic functions.
For $x\in ]-1,1[$ the propagator can be summed up to yield
\bea 
\Delta_{AF}(x,\phi)
= {e^{-i\phi x} \over 2 \cos {\phi\over 2}}
\Big [ e^{i{\phi\over 2}}\theta(x)  - e^{-i{\phi\over 2}}\theta(-x)\Big ] \ .
\eea
For coinciding points ($\tau =\sigma$ i.e. $x=0$) it 
takes the regulated value
\be
\Delta_{AF}(0,\phi) = {i\over 2} \tan {\phi \over 2}
\label{A.8}
\ee
which can be computed by ``symmetric integration'', i.e. symmetrically 
combining the modes $+ r$ and $-r $ and then summing up the series.
Note also that for $x\neq 0$ 
\be
\Delta_{AF}(x,\phi)\Delta_{AF}(-x,\phi) = -{1\over 4} \cos^{-2}{\phi\over 2}
\ee
which, when combined with (\ref{A.8}), shows that this function
has a discontinuity at $x=0$, so that when multiplied by a distribution
it necessitates a regularization.
An example will be discussed at the end of next section.

Let us now review the calculation of the fermionic determinant
\bea
\int_{\mbox{\small{ABC}}} D\bar{\psi}D\psi\ e^{-S} = 
{\det}^D(\partial_{\tau}+i\phi) 
\eea
where ABC stands for antiperiodic boundary conditions and the action
is the one in (\ref{A.4}). 
The easiest way to obtain the determinant is to use the operator 
formalism
\bea
\int_{\mbox{\small{ABC}}} D\bar{\psi}D\psi\ e^{-S}=
{\rm Tr}\, e^{-\hat H_\phi}
\eea
where $\hat H_\phi$ is the hamiltonian operator of the system 
and equals
\bea
\hat H_\phi =i\phi {1\over 2}
(\hat \psi^\dagger_a \hat{\psi}^a 
-\hat{\psi}^a  \hat \psi^\dagger_a )
=i\phi \Big (\hat \psi^\dagger_a \hat{\psi}^a -{D\over 2}\Big ).
\eea 
This is the hamiltonian for a $D$ dimensional fermionic oscillator, 
and the trace is easily computed.
In one dimension the eigenvalues of the fermionic number operator
$\hat \psi^\dagger \hat \psi$ are either 0 or 1, thus one gets
\bea
{\det}^D(\partial_{\tau}+i\phi)&=& {\rm Tr}\, 
e^{-i\phi (\hat \psi^\dagger_a \hat{\psi}^a -{D\over 2})} \nonumber \\
&=& e^{i\phi {D\over 2}}(1+e^{-i\phi})^D = 
\bigg{(}2 \, \mbox{cos} {\phi \over 2}\bigg{)}^D \ .
\eea
Alternatively, one can compute the determinant directly from 
the path integral by expanding the fermions in antiperiodic modes 
and taking the infinite product of eigenvalues
\bea
{\det}^D(\partial_{\tau}+i\phi)&=&
{\det}^D(\partial_{\tau})
\Big [{{\det}(\partial_{\tau}+i\phi)\over 
{\det}(\partial_{\tau})} \Big ]^D 
\nonumber \\
&=&
2^D \prod_{n=-\infty}^{+\infty} \bigg{(} 1+ 
{\phi \over 2\pi(n+{1 \over 2})} \bigg{)}^D 
\nonumber \\
&=&
 \bigg{(}2\, \mbox{cos} {\phi \over 2}\bigg{)}^D
\eea
where we have used a standard representation of the cosine as an 
infinite product (after combining positive and negative frequencies
together as part of our regularization prescription).

\subsection{Dimensional regularization}
\label{Dimensional regularization}

We discuss here the dimensional regularization (DR) of the $N=2$ 
supersymmetric nonlinear sigma models on a compact one-dimensional 
base space, and with target space of arbitrary dimensions,
extending the treatment presented in \cite{Bastianelli:2000nm} for bosonic 
and in \cite{Bastianelli:2002qw} for $N=1$ supersymmetric sigma models.
In particular, we wish to fix the corresponding counterterm.
Dimensional regularization had been previously applied to bosonic
nonlinear sigma models on an infinite one-dimensional base space
and one-dimensional target space in \cite{Kleinert:1999aq},
where it was understood that noncovariant counterterms did not arise,
and in \cite{Bastianelli:2000pt}, where the correct counterterm for 
higher dimensional target space was found.
A general discussion on the regularization issues of one dimensional
nonlinear sigma models can be found in the 
forthcoming book \cite{book}.

The quantum hamiltonian for bosonic systems can be required
to be proportional to the covariant scalar laplacian, 
$H=-{1\over 2}\nabla^2$,
without any coupling to the scalar curvature $R$.
Then the rules of dimensional regularization
developed in \cite{Bastianelli:2000nm} demand the use of a 
counterterm $ V_{DR}= -{1\over 8} R$  to be added to the classical
euclidean action, normalized as $\Delta S_{DR} ={1\over \beta}
\int^1_0 d \tau\, \beta^2 V_{DR}$ (the powers of $\beta$
signal that this is due to a two-loop effect). On the other hand, 
the quantum hamiltonian of the $N=1$ model acts on a spinor space 
and it is fixed by susy to be the square of the Dirac operator
(the susy charge) $H = -{1\over  2}\rldd \rldd 
= -{1\over  2}  (\nabla^2 - {1\over 4} R) $.
In such a case the total counterterm in dimensional regularization
vanish, showing that DR respects $N=1$ susy \cite{Bastianelli:2002qw}.

For the $N=2$ model one might conjecture that the counterterm 
would vanish as well. This is correct, and can be proved in various ways.
One way is to use an arbitrary counterterm proportional to $R$, and then 
fix the proportionality constant so to reproduce some known result, 
e.g. the $a_1$ coefficient for a 0-form.
Then DR is seen to require a vanishing total counterterm.
However, to obtain this result, see (\ref{4.20}), 
one has to integrate over the $U(1)$ modulus $\phi$, 
and one might feel uneasy as the 
modular integration acts effectively as a projection.
A different test which does not require the modular integration
is to compare DR with the unambiguous 
operatorial treatment \cite{Peeters:1993vu}
or with the time slicing regularization scheme worked out 
in \cite{deBoer:1995cb}.
In those papers one finds the transition amplitude for the $N=2$
model. We have used that result to compute the 
partition function (the trace of the transition amplitude)
which is directly related to our path integral on the torus with 
$\phi =0$ and antiperiodic boundary conditions for the fermions.
The comparison then requires the vanishing of the total counterterm
for the $N=2$ model.
Counterterms are local effects arising from ultraviolet ambiguities
and should not depend on the boundary conditions imposed 
on the quantum fields.
The conclusion is that the counterterm vanishes for any value of the 
modular parameter $\phi$.

Dimensional regularization requires the extension
of the space $I=[0,1]$ to $I\times {\mathbb R}^d$.
Then also the action can be extended to $d+1$ dimensions,
so that one can recognize the structure of the vertices and propagators
to dimensionally continue the various Feynman diagrams.
The extended action is 
\bea
 S \eqa {1\over \beta}
\int_{I\times {\mathbb R}^d} 
\! d^{d+1}t \, \Big [{1\over 2}\, g_{\mu\nu}
\partial^\alpha x^\mu \partial_\alpha x^\nu 
+ \bar \psi_a \gamma^\alpha (\partial_\alpha \psi^a 
+ \partial_\alpha  x^\mu \omega_\mu{}^{ab} \psi_b ) 
+i\phi \bar \psi_a \psi^a 
\ccr
&-& {1\over 2} R_{abcd} \bar\psi^a \psi^b \bar \psi^c \psi^d \Big ] 
\eea
where $t^\alpha \equiv (\tau, {\bf t})$ are coordinates in the extended space 
(bold face indicates  vectors in the extra $d$ dimensions)
and $\gamma^\alpha$ are the corresponding Dirac matrices.
The propagators in $d+1$ dimensions read 
\bea 
\Delta(t,s) &=& 
\int {d^d{\bf k}\over (2\pi)^d} \sum_{m=1}^\infty 
{-2\over (\pi m)^2+{\bf k}^2}\,
{\rm sin}(\pi m\tau)\, {\rm sin}(\pi m\sigma)\,
{\rm e}^{i{\bf k}\cdot ({\bf t}-{\bf s})}  
\\[2mm]
\Delta_{gh}(t,s)
&=& \int {d^d{\bf k}\over (2\pi)^d} \sum_{m=1}^\infty 
2\, {\rm sin}(\pi m\tau)\, {\rm sin}(\pi m\sigma)\,
{\rm e}^{i{\bf k}\cdot ({\bf t}-{\bf s})} 
\\ 
&=& 
 \delta (\tau, \sigma)\, \delta^d ({\bf t} -{\bf s}) 
\nonumber  
\\[2mm]
\Delta_{AF}(t-s)
\eqa
 - i \int {d^d{\bf k}\over (2\pi)^d} 
\sum_{r\in Z+{1\over 2}}  { 2 \pi r \gamma^0 
+ {\bf k}\cdot \vec \gamma -\phi   
\over (2 \pi r)^2+{\bf k}^2 -\phi^2}\,
 {\rm e}^{2  \pi i r (\tau-\sigma)}
{\rm e}^{i{\bf k}\cdot ({\bf t}-{\bf s})} 
\eea
and satisfy 
\bea
\partial^\alpha\partial_\alpha \Delta(t,s) \eqa \Delta_{gh}(t,s)=
 \delta (\tau, \sigma)\, \delta^d ({\bf t} -{\bf s}) 
\label{A.18}
\\ 
\Big ( \gamma^\alpha {\partial\over \partial t^\alpha} 
+ i\phi\Big) \Delta_{AF}(t-s)\eqa
\Delta_{AF}(t-s) \Big (- \gamma^\beta {\overleftarrow{
\partial}\over \partial s^\beta} + i \phi\Big )
=  \delta_{AF} (\tau -\sigma)\, \delta^d ({\bf t} -{\bf s})  \ .
\nonumber
\eea
The index contractions in $d+1$ dimensions serve mostly
as a bookkeeping device to keep track of which derivative
can be contracted to which vertex to produce the 
$(d+1)$-dimensional delta function.
The delta functions in (\ref{A.18}) are only to be used in $d+1$ dimensions,
as we assume that only in such a situation the regularization due to 
the extra dimensions is taking place.
Then, by using partial integration one casts the various loop 
integrals in a form which can be computed by sending 
$d\to 0$ first. At this stage one can use the propagators 
in one dimensions, and $\gamma^0=1$ (with no extra factors 
arising from the Dirac algebra in $d+1$ dimensions).  

This procedure will be now exemplified in one of the most relevant graphs 
coming from the vertex 
\bea
\dot x^\mu \omega_{\mu ab}(x) \bpsi^a \psi^b
\quad \leadsto \quad
\partial_\alpha x^\mu \omega_{\mu ab}(x) \bpsi^a \gamma^\alpha\psi^b 
\eea
(we indicate also the extension to $d+1$ dimensions).
Expanding the vertex around $x_0^\mu =x^\mu(\tau) -y^\mu(\tau)$,
and using Riemann normal coordinates and a 
Lorentz gauge such that $\omega_{\mu ab}(x_0)=0$
and $\partial_{(\nu} \omega_{\mu) ab}(x_0)=0$,
one obtains a quartic vertex of the form
\bea
\Delta S \eqa  {1\over \beta} \int_0^1 \!\! d\tau \,  {1\over 2} \,
\dot y^\mu y^\nu R_{\nu\mu ab}(x_0)\bpsi^a \psi^b 
\ccr
& \leadsto & 
{1\over \beta}  \int d^{d+1}t \, {1\over 2} \,
\partial_\alpha y^\mu y^\nu R_{\nu\mu ab}(x_0)\bpsi^a \gamma^\alpha \psi^b \ .
\eea
This vertex can be used to construct the graph 
\bea
&& {1\over 2} \la (\Delta S_3)^2  \ra =
\raisebox{-.78cm}{\scalebox{.7}{ 
{\includegraphics*[149pt,655pt][197pt,720pt]{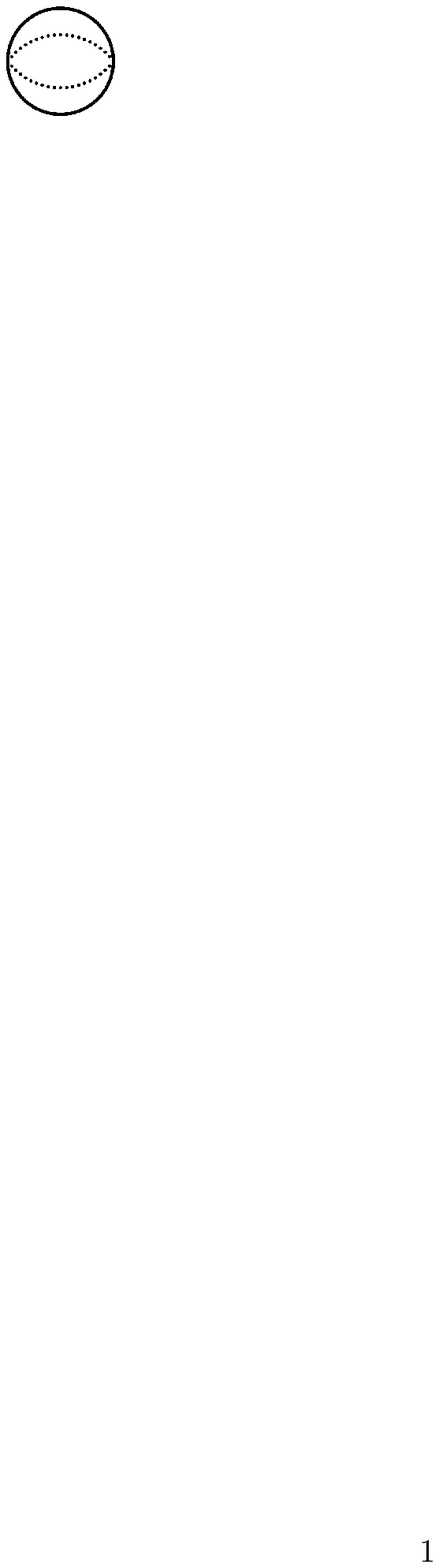}}}}
 = 
{\beta^2\over 8} R_{\mu\nu ab}^2(x_0) I 
\ccr
&& I = 
\int_0^1 \!\!  d\tau  \!\! \int_0^1 \!\!  d\sigma \   
[\ddeld (\tau, \sigma) 
\del (\tau, \sigma)  - 
\ddel (\tau, \sigma) \deld (\tau, \sigma)]  
\Delta_{AF} (\tau-\sigma) \Delta_{AF} (\sigma-\tau)  \ccr
&& \ \ \leadsto  
\int d^{d+1}t \int d^{d+1}s\, 
\Big \{ [{ _\alpha{\Delta_\beta}(t,s)} \del (t,s)  - 
{ _\alpha{\Delta}(t,s)} { {\Delta_\beta}(t,s)}]\ccr
&&\quad \quad \times  {\rm tr}\, [\gamma^\alpha \Delta_{AF} (t-s) 
\gamma^\beta \Delta_{AF} (s-t)] \Big \} 
\label{A.21} 
\eea
where dots and indices on the left/right of the propagators
indicate derivatives with respect to the first/second variable.
Regularization is needed as $ \ddeld $ contains a delta function 
multiplying the step functions contained in $\Delta_{AF}$, and these 
products of distributions are ambiguous and must be carefully regularized.
Thus we have extended the integrals in (\ref{A.21}) to $d+1$ dimensions. 
In DR we can use partial integrations to obtain the relations (\ref{A.18}) 
and we can enforce the delta functions at the regulated level.
In the present case we proceed as follows. 
We integrate by part the $\partial_\alpha$ from 
${ _\alpha{\Delta_\beta}}$. This produces a boundary term which vanish,
a term which doubles the other term in
(\ref{A.21}), and the following extra term
\bea
 && 
\int d^{d+1} t \int d^{d+1}s \,\Big \{ 
\Delta_\beta (t,s) \del (t,s) \,  
 {\rm tr}\, \Big [ \Big ( 
\gamma^\alpha {\partial\over \partial t^\alpha} 
\Delta_{AF} (t,s)\Big ) \gamma^\beta \Delta_{AF} (s,t)
\ccr && \quad\quad\qquad\qquad\qquad
+ \Delta_{AF} (t,s) \gamma^\beta \Big ( \Delta_{AF} (s,t)
{\overleftarrow{ \partial}\over \partial t^\alpha} \gamma^\alpha 
\Big) \Big ]\Big \} \ .
\eea
The ``mass'' term $i\phi$ can be added for free to obtain the Dirac 
equations, then using the second line in (\ref{A.18}) and enforcing
the delta function shows that this extra contribution vanish
\bea
&& 2\int d^{d+1}t \  
\Delta_\beta (t,t) \del (t,t)  
\; {\rm tr}\, [ \gamma^\beta \Delta_{AF} (0)]
\ccr
&& 
\leadsto  \ 
 2 \Delta_{AF} (0) 
\int_0^1 \!\!  d\tau \, \deld (\tau, \tau) \del (\tau, \tau)  =0 \ .
\eea
We have used 
$\deld (\tau, \tau) = \tau - {1\over 2}$ and 
$\del (\tau, \tau)  =\tau^2 -\tau$.
Note that we have removed the regularization only when it was obvious 
that the integral did not contain any dangerous product of distributions 
at $d=0$. Thus, we are left with
\bea 
I\eqa -2 \int d^{d+1}t \int d^{d+1}s \
 { _\alpha{\Delta}(t,s)} { {\Delta_\beta}(t,s)}
\, {\rm tr}\, [\gamma^\alpha \Delta_{AF} (t-s) 
\gamma^\beta \Delta_{AF} (s-t)] \ccr
& \leadsto &
 - 2 \int_0^1 \!\!  d\tau  \!\! \int_0^1 \!\!  d\sigma \   
\ddel (\tau, \sigma) \deld (\tau, \sigma)
\Delta_{AF} (\tau-\sigma) \Delta_{AF} (\sigma-\tau)  
\ccr 
\eqa 
-{1\over 24} \cos^{-2} {\phi\over 2} \ .
\eea
Here we have used that $\ddel (\tau, \sigma) = \sigma -\theta(\sigma -\tau)$,
$\deld (\tau, \sigma) =\tau -\theta(\tau-\sigma)$,
and 
$\Delta_{AF} (\tau-\sigma) \Delta_{AF} (\sigma- \tau)=   
- {1\over 4} \cos^{-2}{\phi\over 2}$
(which are the correct limits of the Fourier sums up to a set of points 
of zero measure). Thus we have obtained 
\be
{1\over 2} \la (\Delta S_3)^2  \ra =
 -{\beta^2\over 192} R^2_{\mu\nu a b} \cos^{-2} {\phi\over 2}
\ee
which is one of the contributions appearing in (\ref{4.10}),
and in fact the only one containing fermions and needing a
regularization. However DR is still needed in other purely bosonic 
graphs.


\end{document}